\documentclass[twocolumn,preprintnumbers,aps]{revtex4}
\usepackage{amsmath}
\usepackage{dcolumn}
\usepackage{bm}
\usepackage{graphicx}
\usepackage{amsfonts}
\usepackage{amssymb}
\usepackage{bbm}
\usepackage{float}

\begin{document}

\title{Quantum speed limit of a photon under non-Markovian dynamics}
\author{Zhen-Yu Xu $^{1}$}
\email{zhenyuxu@suda.edu.cn}
\author{Shi-Qun Zhu $^{1}$}
\email{szhu@suda.edu.cn}
\affiliation{$^{1}$School of Physical Science and Technology, Soochow University, Suzhou,
Jiangsu 215006, China}

\begin{abstract}
Quantum speed limit (QSL) under noise has drawn considerable attention in
real quantum computational processes and quantum communication. Though
non-Markovian noise is proven to be able to accelerate quantum evolution for
a damped Jaynes-Cummings model, in this work we show that non-Markovianity
may even slow down the quantum evolution of an experimentally controllable
photon system. As an important application, QSL time of a photon can be well
controlled by regulating the relevant environment parameter properly, which
is close to reach the currently available photonic experimental technology.
\end{abstract}

\maketitle


\section{Introduction}

A quantum version of \textit{brachistochrone problem} that how fast a
quantum system can evolve between two distinguishable states is of paramount
importance in quantum information processing, for a transition from a state
to its orthogonal one is regarded as the elementary step of a computational
process \cite{book QIC,js}. During the past decades, the study on the
minimum time a quantum state required for reaching its orthogonal one, i.e.,
the quantum speed limit (QSL) time, has been mainly focused on closed
quantum systems with unitary evolution, and a unified lower bound of QSL was
obtained \cite{MT,z1,z2,z3,ML,unified}: $\tau _{\text{QSL}}=\max \left\{ \pi
\hbar /(2\Delta E),\pi \hbar /(2E)\right\} $, where the first quantity in
braces is known as Mandelstam-Tamm (MT) type bound with $\left( \Delta
E\right) ^{2}=\left\langle H^{2}\right\rangle -\left\langle H\right\rangle
^{2}$ and $H$ is the Hamiltonian of the quantum system \cite{MT,z1,z2,z3}
while the second one is referred to as Margolus-Levitin (ML) type bound with
$E=\left\langle H\right\rangle $ \cite{ML}. These bounds, providing a
fundamental limit of the operation rate, are applicable to considerable
quantum tasks such as quantum state transfer \cite{state transfer}, quantum
optimal control \cite{optimal control}, and quantum metrology \cite%
{metrology} and have been extended to nonorthogonal state cases \cite%
{fz1,fz2,fz3,fz4,fz5} and derived from geometric aspects \cite{geo1,geo2}.

In realistic physical processes, however, the quantum systems are open, and
the environmental influence must be taken into account \cite{Book-Open}.
Recently, the QSL time has been extended to nonunitary evolution of open
systems \cite{op1,op2,op3}. Two MT type bounds of QSL, based on the variance
of the generator of the dynamics, were derived and applied to several
typical noisy channels \cite{op1} and estimate the speed limits for quantum
metrology under noise \cite{op2}. Importantly, a unified QSL bound including
both MT and ML types for non-Markovian processes has been introduced in Ref.
\cite{op3}, where the ML bound is also proven to be sharper than the MT
bound. Interestingly, it is discovered that the non-Markovian effect can
speed up the quantum evolution with a damped Jaynes-Cummings (JC) model \cite%
{op3}.

All-optical system has been regarded as an excellent test bed to explore the
foundations of quantum physics as well as quantum information processing
\cite{photon1,photon2}. In this paper, with a photon in a simulated
non-Markovian environment, we show that non-Markovian effect can slow down
the quantum evolution which is contrary to former situation that
non-Markovianity will lead to smaller QSL time for JC model \cite%
{op3}. In addition, we illustrate that the QSL time of a photon can be well
controlled by adjusting the environment parameter. The above phenomena can
be immediately tested with the experimental setups in Refs. \cite%
{xjs,Liu,Tang}.

\begin{figure}[tbp]
\centering
\includegraphics[width=3.5in]{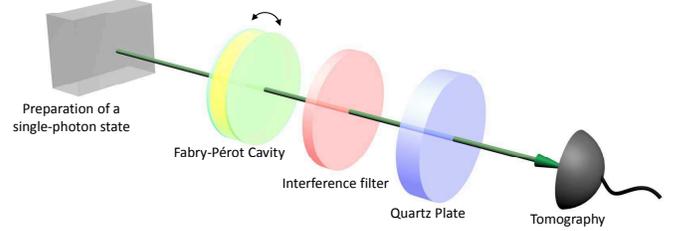}
\caption{A schematic diagram (simplified) of the photonic experimental setup
for testing the non-Markovian effect on quantum speed limit. }
\end{figure}

\section{Non-Markovian model}

The open system we consider in this paper is the polarization degree of a
photon with its frequency functioning as the environment. To simulate the
non-Markovian dynamics of the photon, we employ an experimental setup
containing a rotatable Fabry-P\'{e}rot (FP) cavity followed by an
interference filter and a quartz plate \cite{xjs,Liu,Tang} (see in Fig. 1).
A frequency comb of the photon is generated by a FP cavity and then two
peaks are filtered out through the interference filter. The filtered
frequency distribution $f(\omega )$, representing the probability density of
finding photon in a mode with frequency $\omega $ in this letter, is set to
be a two-peaked Gaussian distribution \cite{Book-optics} $\ $
\begin{equation}
f(\omega )=\frac{\cos ^{2}\xi }{\sqrt{2\pi }\sigma }e^{-\frac{(\omega
-\omega _{1})^{2}}{2\sigma ^{2}}}+\frac{\sin ^{2}\xi }{\sqrt{2\pi }\sigma }%
e^{-\frac{(\omega -\omega _{2})^{2}}{2\sigma ^{2}}},  \label{fw}
\end{equation}%
which are centered at $\omega _{1}$ and $\omega _{2}$ with the same width $%
\sigma $ and is normalized as $\int f(\omega )d\omega =1$. Parameter $\xi
\in \lbrack 0,\pi /2]$ in Eq. (\ref{fw}) controls the relative weight of the
two peaks, which can be adjusted by changing the tilted angle of the FP
cavity \cite{Liu}. Then the non-Markovian dephasing process of the
polarization degree of the photon can occur in the interaction with its
frequency degree in a quartz plate with the following Hamiltonian \cite%
{Hamiltonian}
\begin{equation}
H^{se}=-\hbar \int \left( n_{H}\left\vert H\right\rangle _{s}\left\langle
H\right\vert +n_{V}\left\vert V\right\rangle _{s}\left\langle V\right\vert
\right) \otimes \omega \left\vert \omega \right\rangle _{e}\left\langle
\omega \right\vert d\omega ,  \label{Hamit}
\end{equation}%
where $n_{H(V)}$ is the refraction index of photon in the quartz plate, $%
|H(V)\rangle _{s}$ and $|\omega \rangle _{s}$ are the horizontal(vertical)
polarization and frequency states of the photon.

Provided that an initial product photon state is of the form $\rho
^{s}\otimes \rho ^{e}$, where $\rho ^{s}=(\rho _{jk})_{2\times 2}$ ($j,k=V,H$%
) denotes the polarization state serving as the open system, and $\rho
^{e}=\int d\omega d\omega ^{\prime }F(\omega )F^{\ast }(\omega ^{\prime
})\left\vert \omega \right\rangle _{e}\left\langle \omega ^{\prime
}\right\vert $ is the environmental state with $f(\omega )=|F(\omega )|^{2}$%
. The photon polarization state at time $t$ reads
\begin{equation}
\rho _{t}^{s}=\Lambda _{t}\rho ^{s}=\mathrm{tr}_{e}\left\{ U_{t}\rho
^{s}\otimes \rho ^{e}U_{t}^{\dagger }\right\}  \label{maps}
\end{equation}%
with $\Lambda _{t}$ the quantum map and $U_{t}=\exp [-(i/\hbar
)\int_{0}^{t}dt^{^{\prime }}H^{se}].$ The density matrix of the polarization
degree is then explicitly given by
\begin{equation}
\rho _{t}^{s}=\left(
\begin{array}{cc}
\rho _{VV} & \rho _{VH}\kappa _{t} \\
\rho _{HV}\kappa _{t}^{\ast } & \rho _{HH}%
\end{array}%
\right) ,  \label{statet}
\end{equation}%
where $\kappa _{t}=\int f(\omega )e^{i\omega \Delta nt}d\omega $ is the
dephasing rate with $\Delta n=n_{V}-n_{H}$. By the frequency distribution in
Eq. (\ref{fw}), it takes the form
\begin{equation}
\kappa _{t}=e^{-\frac{\sigma ^{2}\Delta n^{2}t^{2}}{2}}\left( e^{i\omega
_{1}\Delta nt}\cos ^{2}\xi +e^{i\omega _{2}\Delta nt}\sin ^{2}\xi \right) .
\label{k}
\end{equation}

\section{Quantum speed limit for a photon}

A unified lower bound including both MT and ML types for the minimal
evolution time between an initial open system state $\rho =\left\vert \psi
_{0}\right\rangle \left\langle \psi _{0}\right\vert $ and its target state $%
\rho _{\tau }$, governed by the master equation $\dot{\rho}_{t}=\mathcal{L}%
_{t}\rho _{t}$ with $\mathcal{L}_{t}$ the positive generator of the
dynamical semigroup $\Lambda _{t}=\exp (\mathcal{L}_{t}t),$ has been derived
\cite{op3}:
\begin{equation}
\tau _{\text{QSL}}=\max \left\{ \tau _{1},\tau _{2},\tau _{\infty }\right\} ,
\label{QSL1}
\end{equation}%
with
\begin{equation}
\tau _{p}=\frac{1}{\Gamma _{\tau }^{p}}\sin ^{2}\left[ L(\rho ,\rho _{\tau })%
\right] ,\text{ }(p=1,2,\infty )
\end{equation}%
where $\Gamma _{\tau }^{p}=\left( 1/\tau \right) \int_{0}^{\tau
}dt\left\Vert \mathcal{L}_{t}\rho _{t}\right\Vert _{p},$ and $\left\Vert
A\right\Vert _{p}=(a_{1}^{p}+\cdots +a_{n}^{p})^{1/p}$ denotes the $p$-norm
of operator $A$, with $a_{1},\cdots ,a_{n}$ the singular values of operator $%
A$. $L(\rho ,\rho _{\tau })=\arccos \sqrt{\left\langle \psi _{0}\right\vert
\rho _{\tau }\left\vert \psi _{0}\right\rangle }$ represents the Bures angle
between initial pure state $\rho =\left\vert \psi _{0}\right\rangle
\left\langle \psi _{0}\right\vert $ and the target state $\rho _{\tau }$.
Note that $\tau _{1}$ and $\tau _{\infty }$ are bounds of ML type derived by
von Neumann trace inequality while the $\tau _{2}$ is a bound of MT type
deduced according to Cauchy-Schwarz inequality.

With above model, for instance, we will evaluate the minimal evolution time
between states $\rho ^{s}$ and $\rho _{\tau }^{s}$, where $\rho ^{s}$ and $%
\rho _{\tau }^{s}$ denote the states of photon entering and leaving the
quartz plate respectively with $\tau $ the actual driving time when the
photon under non-Markovian dephasing. For simplicity, the initial state is
set to be pure of the form $\rho ^{s}=\left\vert \psi _{0}\right\rangle
\left\langle \psi _{0}\right\vert $ with $\left\vert \psi _{0}\right\rangle
=\sin \alpha \left\vert H\right\rangle +\cos \alpha \left\vert
V\right\rangle $. It is convenient to check that the maximum in Eq. (\ref%
{QSL1}) is $\tau _{\infty },$ for $\Gamma _{\tau }^{\infty }=\Gamma _{\tau
}^{1}/2=\Gamma _{\tau }^{2}/\sqrt{2}$. With Eq. (\ref{QSL1}), the QSL time
of a photon can be written as
\begin{equation}
\tau _{\text{QSL}}=\frac{2\tau \sin ^{2}\theta }{\left\vert \sin 2\alpha
\right\vert \int_{0}^{\tau }dt\left\vert \dot{\kappa}_{t}\right\vert }
\label{qsl1}
\end{equation}%
with
\begin{equation}
\theta =\arccos \sqrt{1-\frac{1}{2}\left( 1-\text{Re}\kappa _{\tau }\right)
\sin ^{2}2\alpha }.  \label{cita}
\end{equation}

\section{Non-Markovian effect on quantum speed limit}

In order to study the non-Markovian effect on the QSL time [Eq. (\ref{qsl1}%
)], we will employ two popular measures for non-Markovianity \footnote{%
There have been introduced several other non-Markovianity measures based on
Fisher information \cite{lxm}, fidelity \cite{fidelity}, mutual information
\cite{luo} etc.. They do not coincide in general \cite{c1,c2,c3,c4,c5}
except for a few cases \cite{luo,zeng}.}: the divisibility of quantum maps
\cite{RHP} and the information flow \cite{BLP,BLP1} based methods.

A quantum map $\Lambda =\{\Lambda _{t}\}_{t\in \lbrack 0,\tau ]}$ is
divisible if $\Lambda _{t}=\Lambda _{t,r}\Lambda _{r}$ for all $0\leq r\leq
t $ with $\Lambda _{t,r}$ completely positive. In Ref. \cite{RHP}, $\Lambda $
is regarded as Markovian if it is divisible, which implies that $\left\Vert
(\Lambda _{t+\epsilon ,t}\otimes \mathbf{1})\rho ^{ss^{\prime }}\right\Vert
_{1}=1,\ \epsilon \geq 0,$ where $\rho ^{ss^{\prime }}=\left\vert \Psi
\right\rangle \left\langle \Psi \right\vert $ with $\left\vert \Psi
\right\rangle =(1/\sqrt{d})\sum_{l=1}^{d}\left\vert l\right\rangle
_{s}\left\vert l\right\rangle _{s^{\prime }}$ a maximally correlated pure
state of the $d$-dimensional open system $s$ and an ancillary system $%
s^{\prime }$. The non-Markovianity is then defined as \cite{RHP},
\begin{equation}
\mathcal{N}_{RHP}(\Lambda )=\int_{h_{t}>0}h_{t}dt  \label{RHP}
\end{equation}%
with $h_{t}=\underset{\epsilon \rightarrow 0^{+}}{\lim }\frac{\left\Vert
(\Lambda _{t+\epsilon ,t}\otimes \mathbf{1})\rho ^{ss^{\prime }}\right\Vert
_{1}-1}{\epsilon }.$

The quantum dynamics $\Lambda $ we consider here is given by $\rho
_{t}^{s}=\Lambda _{t}\rho ^{s}$ [Eq. (\ref{maps})]. Note that $\rho
^{ss^{\prime }}=\left\vert \Psi \right\rangle \left\langle \Psi \right\vert $
with $\left\vert \Psi \right\rangle =(\left\vert H\right\rangle
_{s}\left\vert H\right\rangle _{s^{\prime }}+\left\vert V\right\rangle
_{s}\left\vert V\right\rangle _{s^{\prime }})/\sqrt{2}.$ After simple
calculations, we find that for small $\epsilon $, the non-zero eigenvalues
of $(\Lambda _{t+\epsilon ,t}\otimes \mathbf{1})\rho ^{ss^{\prime }}$ are
\begin{equation}
\frac{1}{2}\pm \frac{1}{2}\sqrt{1+\frac{{\dot{\kappa}}_{t}}{\kappa _{t}}%
\epsilon +\frac{{\dot{\kappa}}_{t}^{\ast }}{\kappa _{t}^{\ast }}\epsilon }%
+o(\epsilon ).  \label{Eigenvalues}
\end{equation}%
We then get
\begin{equation}
h_{t}=\left\{
\begin{array}{r}
\partial _{t}\ln (|\kappa _{t}|),\quad \text{ if }\ \partial _{t}|\kappa
_{t}|>0, \\
0,\quad \text{ \ \ \ \ if }\ \partial _{t}|\kappa _{t}|\leq 0.%
\end{array}%
\right.  \label{ht}
\end{equation}%
According to Eq. (\ref{RHP}), we have
\begin{equation}
\mathcal{N}_{RHP}(\Lambda )=\int_{\partial _{t}|\kappa _{t}|>0}\partial
_{t}\ln (|\kappa _{t}|)dt.  \label{ND}
\end{equation}

The second measure for the non-Markovianity is based on the total amount of
information, characterized by trace distance $D(\Lambda _{t}\rho
_{1}^{s},\Lambda _{t}\rho _{2}^{s})=(1/2)\left\Vert \Lambda _{t}\rho
_{1}^{s}-\Lambda _{t}\rho _{2}^{s}\right\Vert _{1}$ of a pair of evolved
quantum states $(\rho _{1}^{s},\rho _{2}^{s})$, flowing back from the
environment. The direction of information flow is in dependence of the
gradient $g_{t}=\partial _{t}D(\Lambda _{t}\rho _{1}^{s},\Lambda _{t}\rho
_{2}^{s})$, with positive gradient indicating information flowing back to
the system. The non-Markovianity is then defined as \cite{BLP,BLP1}
\begin{equation}
\mathcal{N}_{BLP}(\Lambda )=\underset{\rho _{1}^{s},\rho _{2}^{s}}{\max }%
\int_{g_{t}>0}g_{t}dt,  \label{BLP}
\end{equation}%
where the maximization is over all initial state pairs$.$ For single qubit,
the optimal problem is easy to solve \cite{Liu,xzy,hz} and the optimal trace
distance of the evolved states is found to be $D(\Lambda _{t}\rho
_{1}^{s},\Lambda _{t}\rho _{2}^{s})=\left\vert \kappa _{t}\right\vert $ \cite%
{Liu}. Therefore,
\begin{equation}
\mathcal{N}_{BLP}(\Lambda )=\int_{\partial _{t}|\kappa _{t}|>0}\partial
_{t}\left\vert \kappa _{t}\right\vert dt.  \label{N}
\end{equation}

Since $\ln (|\kappa _{t}|)$ in Eq. (\ref{ND}) owns the same monotonicity as $%
|\kappa _{t}|$ in Eq. (\ref{N}), the divisibility of quantum maps and the
information flow based methods are equivalent in this model. Due to the
simplicity form of $\left\vert \kappa _{t}\right\vert ,$ in the following,
we will focus on the information flow measure (we also drop the subscript
index $BLP$ of $\mathcal{N}_{BLP}(\Lambda )$ for convenience).

The non-Markovianity $\mathcal{N}(\Lambda )$ is dependent on the dephasing
duration $\tau $, which is related to the thickness of the quartz plate. As
an illustration, we consider $\tau \in \lbrack \pi /(\Delta \omega \Delta
n),2\pi /(\Delta \omega \Delta n)]$, and the non-Markovianity reads
\begin{equation}
\mathcal{N}(\Lambda )=\left\vert \kappa _{\tau }\right\vert -\left\vert \cos
2\xi \right\vert e^{-\frac{1}{2}\left( \frac{\pi \sigma }{\Delta \omega }%
\right) ^{2}}.  \label{nonMBLP}
\end{equation}%
For a fixed time $\tau $, the non-Markovianity can be adjusted by the
parameter $\xi $ and two critical points of sudden transition between
Markovian and non-Markovian regions are found to be%
\begin{equation}
\xi _{1}=\frac{1}{2}\arccos (-q)\text{ and }\xi _{2}=\frac{1}{2}\arccos (q),
\label{critical}
\end{equation}%
where $q=\sqrt{v}\left\vert \cos \delta \right\vert /\sqrt{u-v\sin
^{2}\delta },$ $u=e^{\sigma ^{2}\Delta n^{2}\tau ^{2}},\ v=e^{\left( \pi
\sigma /\Delta \omega \right) ^{2}},$ and $\delta =\Delta \omega \Delta
n\tau /2.$

\begin{figure}[tbp]
\centering
\includegraphics[width=3.5in]{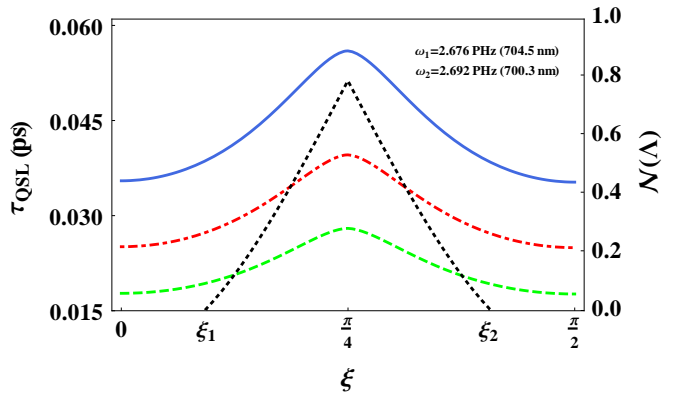}
\caption{Non-Markovian effect on quantum speed limit (QSL) of a photon under
dephasing nosie. QSL time $\protect\tau _{\infty }$ (blue solid curve), $%
\protect\tau _{2}$ (red dot-dashed curve), $\protect\tau _{1}$ (green dashed
curve), and $\mathcal{N}$($\Lambda $) (black dotted curve) as a function of
parameter $\protect\xi $ controlling the relative height of two peaks of
frequency distribution. The initial state with $\protect\alpha =\protect\pi %
/4$ evolves during an actual driving time $\protect\tau =2\protect\pi %
/(\Delta \protect\omega \Delta n)\simeq 0.39$ ps with $\Delta n=0.01$, $%
\protect\sigma =1.8$ THz, $\protect\omega _{1}=2.676$ PHz ($\simeq 704.5$
nm), and $\protect\omega _{2}=2.692$ PHz ($\simeq 700.3$ nm) \protect\cite%
{Liu}.}
\end{figure}

In Fig. 2, the QSL time $\tau _{\text{QSL}}(\tau _{\infty })$ (blue solid
curve) together with $\tau _{2}$ (red dot-dashed curve) and $\tau _{1}$
(green dashed curve) and non-Markovianity $\mathcal{N}(\Lambda )$ (black
dotted curve) are compared to parameter $\xi $ in the case $\alpha =\pi /4$
and $\tau =2\pi /(\Delta \omega \Delta n)$ with $\Delta \omega =\omega
_{2}-\omega _{1}$, where the related parameters are all selected according
to experimental data with $\Delta n=0.01$, $\sigma =1.8$ THz, $\omega
_{1}=2.676$ PHz ($\simeq 704.5$ nm), and $\omega _{2}=2.692$ PHz ($\simeq
700.3$ nm) \cite{Liu}.

The most remarkable feature appeared in Fig. 2 is that the non-Markovian
effect will slow down the quantum evolution, for the monotonicity of $%
\mathcal{N}(\Lambda )$ is in agreement with $\tau _{\text{QSL}}$ in the
non-Markovian region $\xi \in \lbrack \xi _{1},\xi _{2}]$. By controlling
the environment parameter $\xi $ (related to the tilted angle of the FP
cavity)$,$ QSL time can be well controlled. The above phenomenon that the
stronger the non-Markovianity, the longer time required to reach the target
state is just the opposite side illustrated in Ref. \cite{op3}, where the
non-Markovian effect will speed up the evolution for corresponding model. To
our common wisdom, the non-Markovianity reflects the memory effect of the
environment, which is usually thought as beneficial in quantum tasks \cite%
{nonMji}. The above phenomenon, however, implies that for a photon, the
non-Markovian effect may slow down its quantum evolution which may do harm
to quantum computational processes with photon systems \cite{computation}.

\section{Conclusion}

In this paper, with a photonic non-Markovian dephasing model, we illustrate
that the non-Markovian effect can slow down the quantum speed limit, which
presents an opposite effect of non-Markovianity that can speed up the
quantum evolution for a JC model. The phenomenon we illustrated in this work
is analyzed by real experimental data and can be tested immediately by the
all-optical setups in Refs. \cite{xjs,Liu,Tang}. A strict theorem whenever
the non-Markovian effect can speed up or slow down the quantum evolution is
still not clear, however, the answer to this question is of great
importance, especially in quantum computational processes under noise.

Zhen-Yu Xu acknowledges Yu-Li Dong and Fei Zhou for valuable discussions.
This work was supported by NNSFC under Grant Nos. 11204196, 11074184, and
SRFDPHE under Grant. No. 20123201120004.

\end{document}